\documentclass{aa}

\usepackage{graphicx}

\newcommand{\change}{}

\begin{document}

\title{Accurate magnetic field measurements of Vega-like stars and Herbig Ae/Be stars\thanks{Based on observations obtained
at the European Southern Observatory, Paranal, Chile (ESO programme Nos.~072.D-0377, 073.D-0464, 074.C-0463 and 075.D-0507).}}

\author{S. Hubrig\inst{1} \and R.\,V. Yudin\inst{2,3}
\and M. Sch\"oller\inst{1}
\and M.\,A. Pogodin\inst{2,3}}

\institute{European Southern Observatory, Casilla 19001, Santiago 19, Chile
\and Central Astronomical Observatory of the Russian Academy of Sciences at
Pulkovo, 196140 Saint-Petersburg, Russia
\and
Isaac Newton Institute of Chile, St.-Petersburg Branch, Russia
}

\date{Received xx/Accepted yy}

\offprints{S. Hubrig}

\titlerunning{Magnetic fields of Vega-like stars}
\authorrunning{Hubrig et~al.}

\abstract{
We obtained accurate circular spectropolarimetric observations
of a sample of Vega-like and Herbig Ae/Be stars with FORS\,1 at the VLT
in an attempt to detect their magnetic fields.
No magnetic field could be diagnosed in any Vega-like star.
The most accurate determination of a magnetic field, at 2.6\,$\sigma$ level, 
was performed for the Vega-like star $\iota$\,Cen, for which we
measured $\left<B_z\right>$=$-77\pm$30\,G.
In the prototype of Vega-like stars, the star $\beta$~Pictoris,
which shows conspicuous signs of chromospheric activity,
a longitudinal magnetic field is measured only at $\sim$1.5\,$\sigma$ level.
We diagnosed a longitudinal magnetic field for the first time at a level higher than 
3\,$\sigma$ for the two Herbig Ae stars HD\,31648 and HD\,144432 and confirm 
the existence of a previously detected
magnetic field in a third Herbig Ae star, HD\,139614.
Finally, we discuss the discovery of distinctive Zeeman features
in the unusual Herbig Ae star HD\,190073, where the \ion{Ca}{ii} doublet
displays several components in both H and K lines.
From the measurement of circular polarization in all Balmer lines from
H$_\beta$ to H$_8$, we obtain $\left<B_z\right>$=$+26\pm$34\,G.
However, using only the \ion{Ca}{ii} H and K lines
for the measurement of circular polarization, we are able to diagnose
a longitudinal magnetic field at 
2.8\,$\sigma$ level, $\left<B_z\right>$=$+84\pm$30\,G.

           }

\maketitle

\section{Introduction}\label{sec1}

Since the discovery of the large infrared excess at wavelengths longward of
12\,$\mu{}m$ from Vega (Aumann et~al.\ \cite{Au84}),
more than one hundred main-sequence stars have been found to have similar infrared excess in 
the IRAS wavebands and have been called Vega-like stars. 
They are predominantly spectral type A main-sequence stars with continuum 
emission at far-infrared 
wavelengths in excess of that expected from the photospheres of stars of their 
spectral type. The excess flux is ascribed to thermal emission
from orbiting dust grains with a temperature of about 100\,K (Aumann \cite{Au85}). 

 One of the prototypes of Vega-like stars,
$\beta$~Pictoris, shows evidence of a complex stellar environment
with dense and hot regions. It was recently argued that the observed
broad emission from the highly ionised species \ion{C}{iii} and \ion{O}{iv} may
originate from a solar-like extended chromosphere or from
magnetospheric accretion (Deleuil et~al.\ \cite{De01}). In the
theoretical framework, the disk of this star is truncated at some
inner radius, and infalling gas from the disk is
channeled by an axisymmetric dipole magnetic field connecting the
disk to the star (Hartmann et~al.\ \cite{Ha94}).
If solar-like activity is responsible for the
observed highly ionised species, a theoretical model to explain the origin and 
evolution of these phenomena remains to be built. At the moment, the presence
of magnetic fields in Vega-like stars has not been convincingly tested whether observationally
or theoretically.

Many observational properties (e.g.\ spectral energy distribution, size and 
geometrical 
characteristics of the circumstellar (CS) disk, and, possibly, stellar magnetic field) of young 
stars depend 
on the stage of evolution of their CS shells. An evolutionary approach to 
analysing polarimetric data for Herbig Ae/Be and Vega-like stars was recently suggested by 
Yudin (\cite{Yu00}). Herbig Ae/Be stars have spectral types from B to F8 and were first 
defined as a class in \cite{Her60} by Herbig.
They show emission lines
and display a flux excess in the IR due to the presence of cool and/or hot circumstellar
matter (P\'erez \& Grady \cite{PG97}).
It was shown that most young stars have a statistically higher 
value of polarization in comparison to stars that are at more advanced stages of 
evolution on the main sequence (Yudin \cite{Yu00}). The evolutionary approach can probably also 
be applied to 
investigations of magnetic fields in Vega-like stars and Herbig Ae/Be stars.
It is very likely that Herbig Ae/Be stars are the precursors
of most Vega-like stars. The dereddened optical colours of
Vega-like stars are consistent with photospheric emission from
main-sequence stars. However, for a number of Vega-like stars,
the middle/far-IR photometry revealed excess IR emission with colours
resembling those of Herbig Ae/Be systems
(Sylvester et~al.\ \cite{Syl96}; Sheret et~al.\ \cite{Sher04}).

The only attempt to measure a magnetic field in a Vega-like star was
done more than twenty years ago by Landstreet (\cite{Lan82}). Using a
two-channel photoelectric cell polarimeter as a Balmer--line Zeeman analyser,
he failed to detect a longitudinal magnetic field in $\beta$\,Leo with 1$\sigma$
error bars of 65\,G.

Although magnetic fields are believed to play a crucial role in controlling
accretion onto, and winds from, Herbig Ae/Be stars, there is still no 
observational evidence demonstrating the strength, extent, and geometry
of their magnetic fields. 
There were several attempts in the past to measure
magnetic fields in Herbig Ae/Be stars
(e.g., Glagolevskij \& Chountonov \cite{GlCh01}; Catala et~al.\ \cite{Ca93}),
but the first definite evidence of magnetic fields in
Herbig Ae/Be stars has only recently been presented for the star
HD\,139614, where a longitudinal magnetic field at
4.8\,$\sigma$ level, $\left<B_z\right>$=$-$$450\pm93$\,G was diagnosed
(Hubrig et~al.\ \cite{Hu04b}).
For another Herbig Ae star, HD\,104237, a marginal circular polarization
signature indicating a longitudinal magnetic field 
of $\sim$50\,G was observed in metal lines by Donati et~al.\ (\cite{Do97}).

On the one hand, the lack of detections of magnetic fields in Herbig Ae/Be stars could 
plausibly be explained by the low accuracy of previous measurements with typical errors
up to 1000\,G (Glagolevskij \& Chountonov \cite{GlCh01}).
On the other hand, non-detection of magnetic fields in Vega-like and Herbig Ae/Be stars may
indicate that large-scale, organised magnetic fields with strengths in excess of
a few hundred Gauss are not widespread among these stars.
It is obvious that any further attempt to detect fields in these stars must
aim at achieving considerably better accuracy and at applying a 
special observational strategy. The method of Glagolevskij \& Chountonov
(\cite{GlCh01}) allowed them to determine only the average value of the 
magnetic field from the measurements of a large number of metal spectral lines.
However, it is quite possible that, in stars with CS disks, the magnetic field 
has a complex structure and that strong fields exist only locally, but do so with a topology
such that the measurements yield different strength and polarity for different
elements and ions.
As an example, Babcock (\cite{Bab58}) measured a magnetic field of $\sim$$-$270\,G
on \ion{Ca}{ii} H and K components, $\sim$$+$270\,G for lines of neutral
metals, and $\sim$0\,G for lines of ionised metals
for the Herbig Ae star HD\,190073 with a circumstellar disk.

We should note that in our recent study of a few Herbig Ae stars with FORS\,1 at 
the VLT in spectropolarimetric mode (Hubrig et~al.\ \cite{Hu04b}),
we also reported the presence of distinctive circular polarization signatures in the Stokes V spectra of 
the \ion{Ca}{ii} H and K lines, which appear unresolved at the low spectral
resolution achievable with FORS\,1 (R$\sim$2000). 
We also mentioned that, according to a 
previous study of the \ion{Ca}{ii} emissivity from dense envelopes
near Herbig stars, which used photoionization calculations
(Hamann \& Persson \cite{HP92}), it is very likely that
the \ion{Ca}{ii} lines form very near the photosphere.
It is also probable that
models involving accretion of circumstellar matter from the disk to the
star along a global stellar magnetic field of a specific
geometry can account for the Zeeman signatures observed in the \ion{Ca}{ii} lines.
However, the actual geometry of magnetic fields in Herbig Ae stars is 
completely unknown, and further spectropolarimetric observations at higher resolution in the 
spectral region around the \ion{Ca}{ii} H and K lines, as well as other metal lines,
could provide an essential clue to the accretion
mechanism and the geometry of magnetic fields in these stars.

In this paper, we report an attempt to detect a magnetic field in a small sample of 
Vega-like and
Herbig Ae/Be stars by using circular spectropolarimetric observations obtained with FORS\,1 
with GRISM\,1200g at a resolution of R$\sim$4000, which is twice as high as
the one in our previous studies.
Our recent magnetic field measurements
in the broad hydrogen lines of A and B-type stars demonstrated that magnetic fields can be 
measured with an error bar lower than 50\,G (Hubrig et~al.\ \cite{Hu05b}).
Encouraged by these results we decided to carry out subsequent observations
with a non-standard readout mode
with a low gain (A,1x1,low), which allows
the signal-to-noise ratio to be increased by a factor of $\approx$2.
In the following we show that the FORS\,1 instrument in spectropolarimetric mode is well
suited  for accurate magnetic field measurements
of the generally broad-lined Vega-like and 
Herbig Ae/Be stars.

\begin{table*}
\label{tab:basic} \caption{Basic data of the studied Herbig Ae/Be and
Vega-like stars.}
\begin{center}
\begin{tabular}{rlccccrcccc}
\hline\noalign{\smallskip}
\multicolumn{1}{c}{HD} &
\multicolumn{1}{c}{Other} &
\multicolumn{1}{c}{$V$} &
\multicolumn{1}{c}{Sp.\ Type} &
\multicolumn{1}{c}{$\log T_{\rm eff}$} &
\multicolumn{1}{c}{$\log g$} &
\multicolumn{1}{c}{$v\sin{i}$}&
\multicolumn{1}{c}{E(V-L)} &
\multicolumn{1}{c}{$P$ [\%]} &
\multicolumn{1}{c}{Age [Myr]} &
\multicolumn{1}{c}{Ref.}
\\
\hline\noalign{\smallskip} \multicolumn{11}{c}{Herbig Ae/Be stars}\\
\noalign{\smallskip}
31648 & MWC 480    & 7.7   &A3pesh&3.90-3.95& 3.5  &102&2$\fm{}61$&$\sim$0.3 &2.5--6 & 1,2,3\\
38238 & V351 Ori   &8.9 &A7IIIe&3.87-3.89& 3.6-3.68  &$\sim$100 & $2{^m}$--$4{^m}$&-- & 1--3 & 4,5,6,7,8\\
139614& CD-27 10778& 8.2&A7Ve&3.88-3.92& 4.0-4.5&13&2$\fm{}0$&0.15 & 5--10& 6,9,10\\
144432& CD-42 10650&8.2&A9Ve &3.86-3.87& 3.9-4.3&54&1$\fm{}94$&0.1--0.5 &1-3 & 2,9,10,11,12\\
144668&V856 Sco &7.0&A7IVe& 3.89& 3.5-4.0&204&$3{^m}$--$3\fm{}5$&0.5--1.3 &0.5--1& 2,3,6,10\\
163296& MWC 275    & 6.9 &A1Ve&3.97& 4.0&133&2$\fm{}96$&0.5--1.3 &3--5 & 1,2,3,6,10,13\\
190073& MWC 325    & 7.8 &A2IIIe&3.95& 3.5&15&2$\fm{}46$&0.4&3--5 & 1,6,14\\
\hline\noalign{\smallskip}

\multicolumn{11}{c}{Vega-like stars}\\
\noalign{\smallskip}
26676  & HR\,1307              &6.2 &B8Vn&4.06-4.08&4.0 &175&$\sim 0\fm$1&0.59&$\sim$70 & 15,16\\
39060  & $\beta$\,Pic &3.9&A5V &3.91-3.96&4.35 &130&0$\fm{}01$&$\sim$0.02 &20--280& 17,18,19,20\\
61712  & CD-43 3425            &9.0&B7/B8V& -- &-- &{--}&$\sim 0\fm$2&-- &--& \\
102647 & $\beta$\,Leo &2.1&A3V &3.92-3.96&4.30 &128&-0$\fm{}03$&$\sim$0.03 &15--330&17,19,20,22 \\
109085 & $\eta$\,Crv  &4.3&F2V &3.83-3.84&4.26 &92&-0$\fm{}13$&$\sim$0.05 &950--1560 & 11,23,24,25\\
115892 & $\iota$\,Cen &2.7&A2V &3.96-3.97&-- &90&-0$\fm{}14$&$\sim$0.01 &$\sim$500& 25\\
219571 & $\gamma$\,Tuc&4.0&F1III &3.82-3.88&3.87-3.89 &90&0$\fm{}04$&0.005 &1700--5720 & 11,23,24,25\\

\hline\noalign{\smallskip}
\end{tabular}
\end{center}
[1]~Pogodin et~al.\ (\cite{Pog05}),
[2]~Acke \& van den Ancker (\cite{Ack04}),
[3]~Natta et~al.\ (\cite{Nat97}),
[4]~Kovalchuk \& Pugach (\cite{Koval97}),
[5]~Hernandez et~al.\ (\cite{He05}),
[6]~van den Ancker et~al.\ (\cite{Anc98}),
[7]~Marconi et~al.\ (\cite{Mar00}),
[8]~Balona et~al.\ (\cite{Bal02}),
[9]~Dunkin et~al.\ (\cite{Du97}),
[10]~Sartori et~al.\ (\cite{Sar03}),
[11]~Nordstroem et~al.\ (\cite{Nor04}),
[12]~P\'erez et~al.\ (\cite{Per04}),
[13]~Swartz et~al.\ (\cite{Swar05}),
[14]~Corporon \& Lagrange (\cite{Cor99}),
[15]~Stickland (\cite{Stic79}),
[16]~Voshchinnikov \& Marchenko (\cite{Vos82}),
[17]~Song et~al.\ (\cite{So01}),
[18]~Sheret et~al.\ (\cite{Sher04}),
[19]~Lachaume et~al.\ (\cite{Lach99}),
[20]~Habing et~al.\ (\cite{Hab01}),
[21]~Holmes et~al.\ (\cite{Holm03}),
[22]~Metchev et~al.\ (\cite{Met04}),
[23]~Ibukiyama \& Arimito (\cite{Ibu02}),
[24]~Marsakov \& Shevelev (\cite{Mar95}),
[25]~Wyatt et~al.\ (\cite{Wy05}),
[26]~Greaves \& Wyatt (\cite{Gre03}),
[27]~Lambert \& Reddy (\cite{Lamb04}).

\end{table*}

\section{Sample of stars}\label{sec2}

The basic data of the Herbig Ae/Be and Vega-like stars studied here are presented in
Table~\ref{tab:basic}. The columns indicate, in order, the HD number of the star, another
identifier, the visual magnitude, the spectral type, and the stellar
parameters $\log T_{\rm eff}$, $\log g$, and $v\sin{i}$. The last four columns
list the near-infrared colour excess $E(V-L)$, the linear
polarization $P$, estimates of ages of the stars, and corresponding references.
Visual magnitudes, spectral types, and $v\sin{i}$ values were taken
from the SIMBAD database, and the values for the linear polarization and the infrared colour 
excess $E(V-L)$ were retrieved from Yudin\ (\cite{Yu00}) and references therein.
From these data, it is clearly seen that the younger stars have a statistically higher
value of polarization and near-IR excess in comparison to stars
at a more advanced stage of evolution.
One additional normal A-type star, HD\,104231, for which previous longitudinal field measurements
showed no evidence for the presence of magnetic fields
(Shorlin et~al.\ \cite{Sho02}), was selected to check that the
instrument is functioning properly and does not introduce any spurious signals. 
The result of our measurements, $\left<B_z\right>$=$+30\pm$37\,G,
is fully consistent with the previous results of Shorlin et~al.\ (\cite{Sho02})
who obtained $\left<B_z\right>$=$-93\pm$53\,G.


\begin{table}
\label{tab:results} \caption{ Results of magnetic field measurements
of Herbig Ae/Be and Vega-like stars.}
\begin{center}
\begin{tabular}{rlrc}
\hline\noalign{\smallskip} \multicolumn{1}{c}{HD} &
\multicolumn{1}{c}{MJD}  &
\multicolumn{1}{c}{$\left<B_z\right>$} &
\multicolumn{1}{c}{Remarks} \\
\hline\noalign{\smallskip} \multicolumn{4}{c}{Herbig Ae/Be
stars}\\
\noalign{\smallskip}
31648 &53296.35&$+87\pm22$\,G & \\
38238 &53249.37&$-115\pm67$\,G & \\
139614&52904.04&$-450\pm93$\,G & \\
139614&53405.37&$-116\pm34$\,G & \\
144432&52900.99&$-94\pm60$\,G & \\
144432&53447.35&$-119\pm38$\,G & \\
144668&52901.01&$-118\pm48$\,G & \\
144668&53120.25&$-107\pm40$\,G & \\
163296&53279.00&$-57\pm33$\,G & \\
190073&53519.38&$+84\pm30$\,G & \ion{Ca}{ii} H+K \\
\hline\noalign{\smallskip}

\multicolumn{4}{c}{Vega-like stars}\\
\noalign{\smallskip}
26676 &53278.33& $+42\pm136$\,G & \\
39060 &53296.26& $-79\pm53$\,G & \\
61712 &53359.25& $-16\pm47$\,G & \\
102647&53403.37& $-107\pm76$\,G & \\
109085&53405.32& $-114\pm59$\,G & \\
115892&53405.34& $-77\pm30$\,G & \\
219571&53279.03& $-49\pm41$\,G & \\
\hline\noalign{\smallskip}

\multicolumn{4}{c}{Normal A star}\\
\noalign{\smallskip}
104321&53385.37& $+30\pm37$\,G & \\

\hline\noalign{\smallskip}
\end{tabular}
\end{center}

\end{table}

\section{Observations and data reduction}\label{sec3}

\begin{figure*}
\centering
\includegraphics[width=0.30\textwidth]{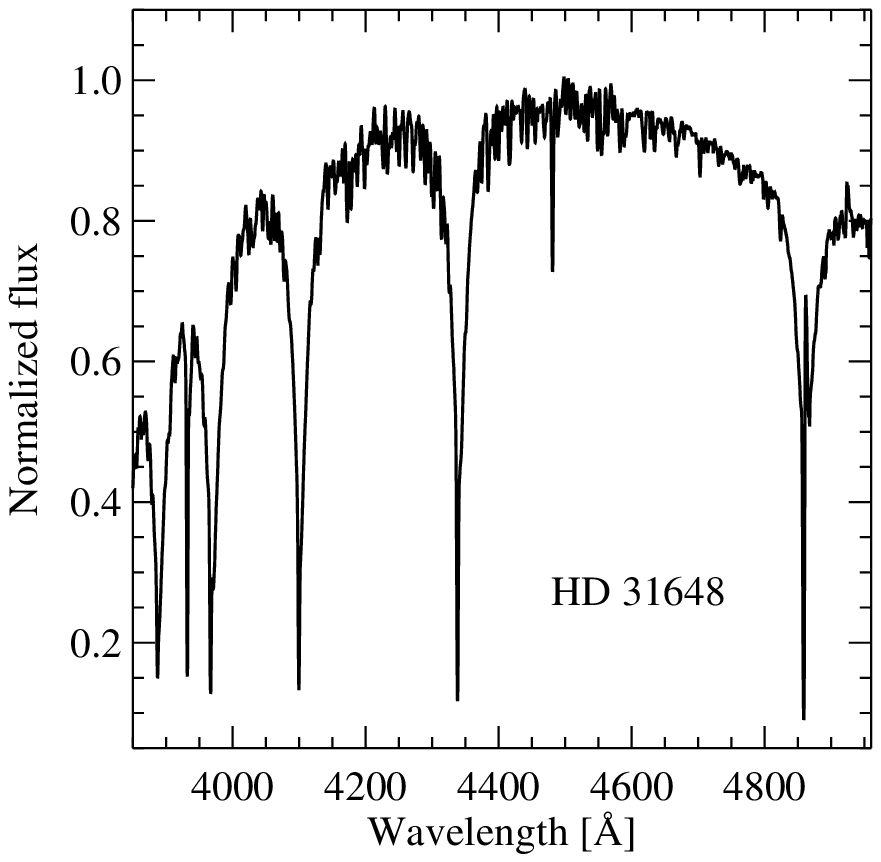}
\includegraphics[width=0.30\textwidth]{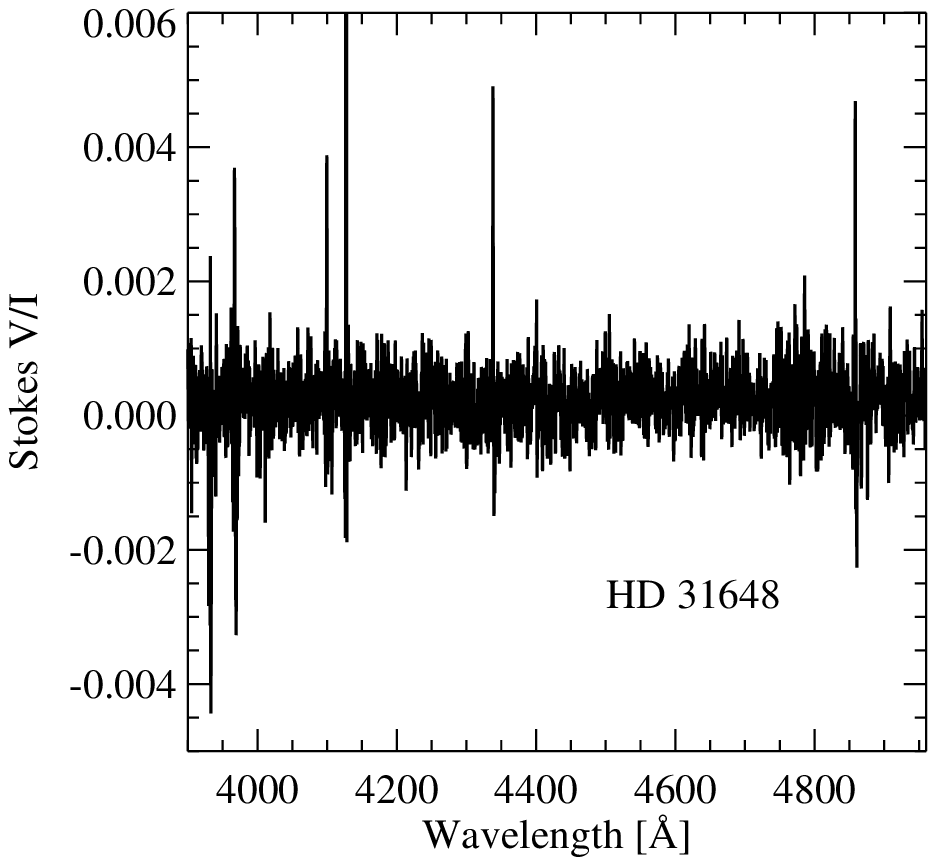}
\includegraphics[width=0.30\textwidth]{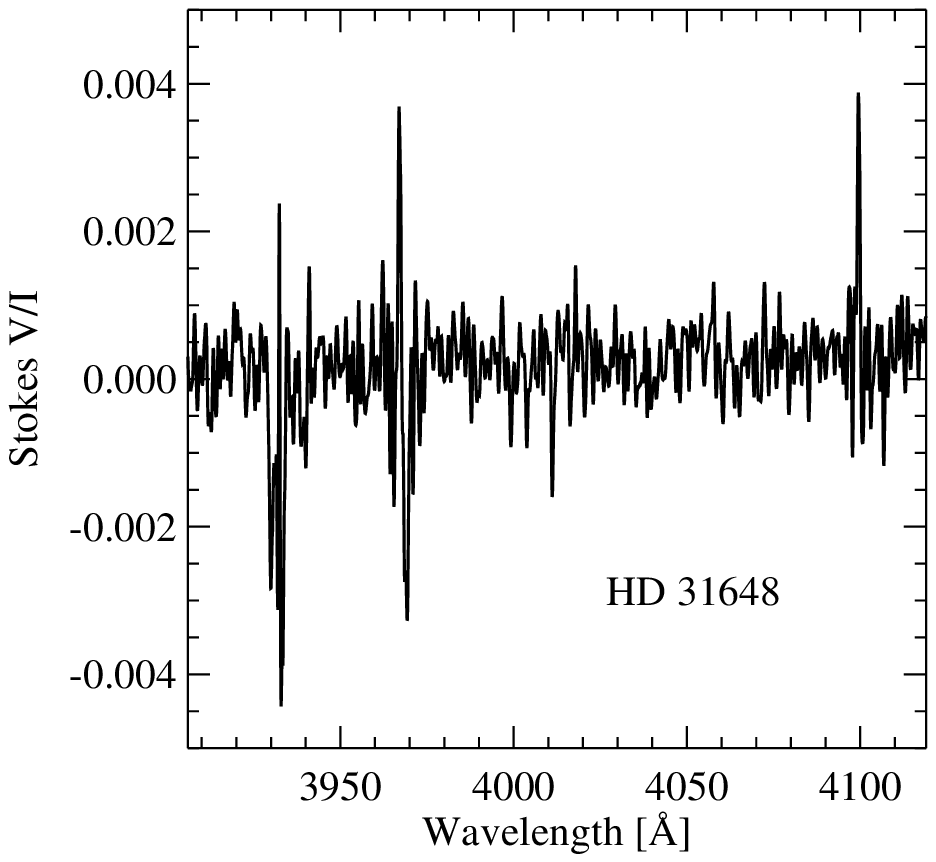}
\caption{
Stokes spectra of the Herbig Ae star HD\,31648:
Normalised Stokes I spectrum (left) and
Stokes V/I spectrum over the full range (centre)
and around the \ion{Ca}{ii} doublet and nearby H Balmer lines (right).
Zeeman signatures are clearly seen at the locations of the \ion{Ca}{ii} H and K
and H Balmer lines.
}
\label{fig:Stokes31648}
\end{figure*}

All longitudinal field determinations for the stars in our sample were
obtained from the observations with FORS\,1 at the VLT in service
mode from April 2003 to June 2005.
The multi-mode instrument
FORS\,1 is equipped with polarization analyzing optics comprising
super-achromatic half-wave and quarter-wave phase retarder plates
and a Wollaston prism with a beam divergence of
22$^{\prime\prime}$ in standard resolution mode. We used the
GRISM\,1200g to cover the H Balmer lines from H$_\beta$ to H$_8$,
and the narrowest available slit width of 0$\farcs$4 to obtain a
spectral resolving power of R$\sim$4000. For two Herbig Ae/Be stars,
HD\,38238 and HD\,144668, 
we used the GRISM\,600B to cover all H  Balmer lines from H$_\beta$ to the Balmer jump.
The spectral resolution of the FORS\,1 spectra of these stars is
about 2000.
 
Assessment of the
longitudinal magnetic field using the FORS\,1 spectra is achieved
by measuring the circular polarization of opposite sign induced in
the wings of broad lines, such as Balmer lines, by the Zeeman
effect. Measurement of circular polarization in magnetically
sensitive lines is the most direct means of detecting magnetic
fields on stellar surfaces. The errors in the measurements of the
polarization were determined from photon counting statistics
and converted to errors of field measurements. 
{\change To obtain a very large signal-to-noise ratio that corresponds to 
mean longitudinal magnetic field error bars lower than 50\,G, we took
four to eight continuous series of two exposures
for each star in our sample using a non-standard readout mode
with low gain (A,1x1,low) and with the retarder waveplate oriented at two different
angles, $+$45$^\circ$ and $-$45$^\circ$.
In comparison to our previous observations with FORS\,1 with the standard readout mode
(A,1x1,high), use of the low gain allowed to increase the signal-to-noise ratio
of our spectra by a factor of $\approx$2. The Zeeman effect increases as
$\lambda^2$, whereas other line-broadening effects depend linearly on $\lambda$.
Hence, as we mainly used the GRISM\,1200g,
the error associated with determining the mean longitudinal magnetic field from 
Balmer lines at longer wavelengths is smaller than when obtained from Balmer lines at shorter wavelengths.
The magnetic sensitivity is further enhanced by adding the signal of all individual
Balmer lines from H$_\beta$ to H$_8$ that are observed simultaneously with this setup.
In the end}, the spectropolarimetric capability of the FORS\,1
instrument in combination with the large light collecting power of
the VLT allowed us to achieve a S/N ratio of up to a few thousand
per pixel around 4600\,\AA{} in the one-dimensional spectrum.
Only for one Vega-like star, HD\,26676, is the signal-to-noise ratio about a few hundreds
due to mediocre weather conditions during the observing night.
More details on the observing technique with FORS\,1
can be found elsewhere (e.g., Bagnulo et~al.\ \cite{Ba02};
Hubrig et~al.\ \cite{Hu04a}).

\section{Results}\label{sec4}

\begin{figure*}
\centering
\includegraphics[width=0.30\textwidth]{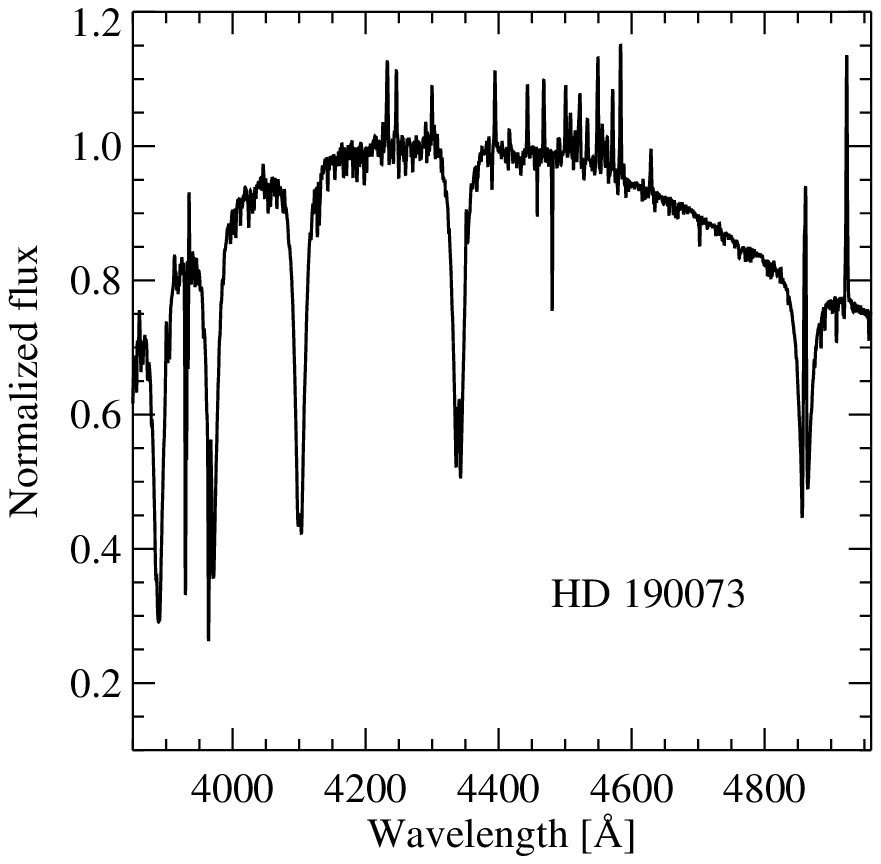}
\includegraphics[width=0.30\textwidth]{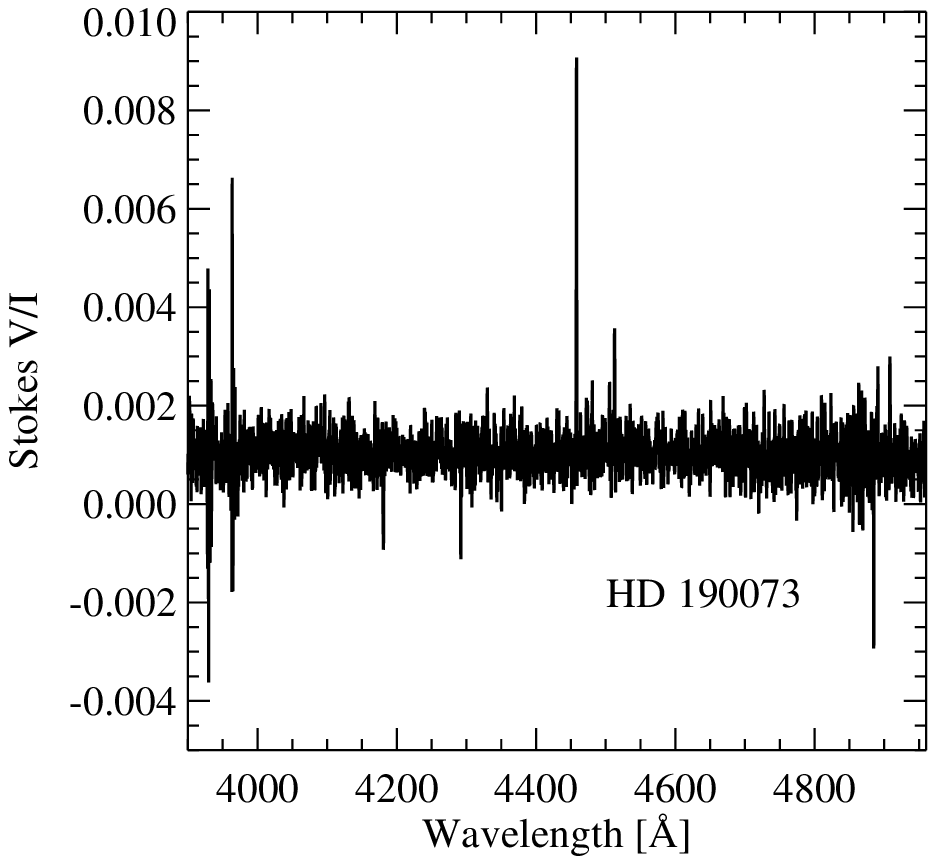}
\includegraphics[width=0.30\textwidth]{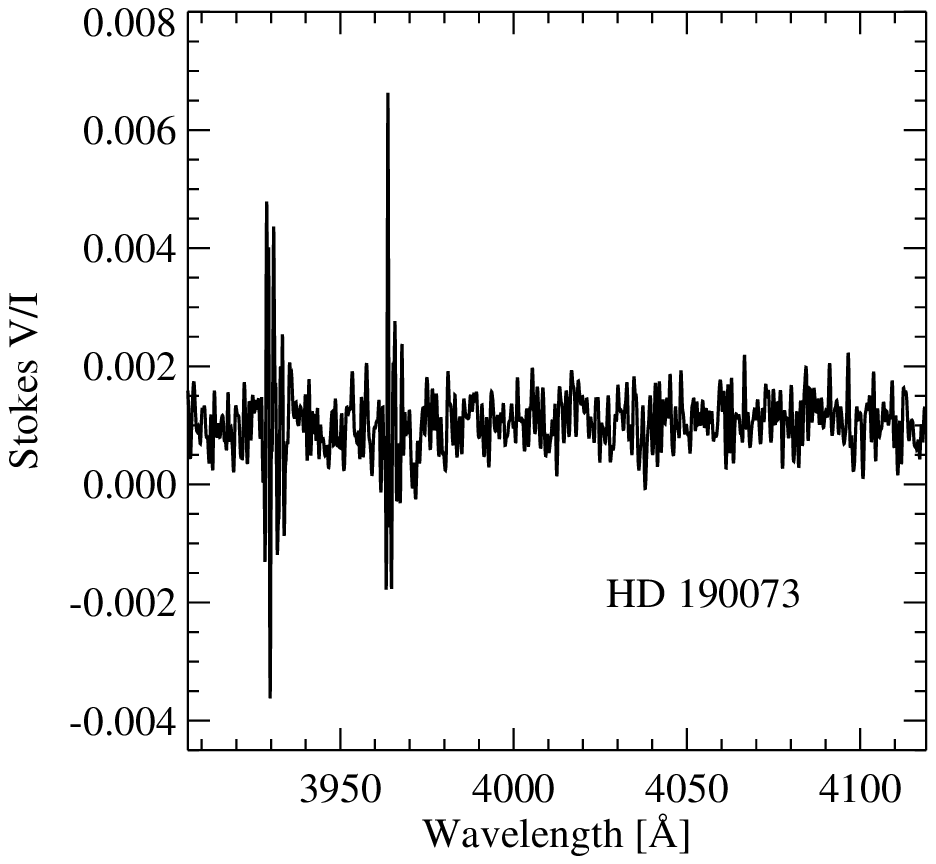}
\caption{
Stokes spectra of the Herbig Ae star HD\,190073:
Normalised Stokes I spectrum (left) and
Stokes V/I spectrum over the full range (centre)
and around the \ion{Ca}{ii} doublet and nearby H Balmer lines (right).
The spectrum is extremely rich in emission lines demonstrating the
presence of a strong stellar wind and a dense circumstellar disk.
Zeeman signatures are clearly seen in the \ion{Ca}{ii} H and K lines,
but they are not detectable at the locations of the H Balmer lines.
}
\label{fig:Stokes190073}
\end{figure*}

We present in Table~\ref{tab:results} our determination of the mean
longitudinal magnetic field $\left<B_z\right>$ for all
stars of both samples with the corresponding Modified Julian Dates of
the observations. For completeness, our previous measurements of three Herbig Ae stars,
HD\,139614, HD\,144432, and HD\,144668, are also included.
The mean longitudinal magnetic field
is the average over the stellar hemisphere visible at the time of
observation of the component of the magnetic field parallel to the
line of sight, weighted by the local emergent spectral line
intensity. It is diagnosed from the slope of a linear regression of
$V/I$ versus the quantity $-g_{\rm eff} \Delta\lambda_z \lambda^2
\frac{1}{I} \frac{{\mathrm d}I}{{\mathrm d}\lambda} \left<B_z\right
> + V_0/I_0$.
This procedure is described in detail by
Bagnulo et~al.\ (\cite{Ba02}) and Hubrig et~al.\ (\cite{Hu04a}).
Our experience from a study of a large sample of magnetic and
non-magnetic Ap and Bp stars revealed  that this regression
technique is very robust and that detections with
$B_z > 3\,\sigma$
result only for stars possessing magnetic fields
(Hubrig et~al.\ \cite{Hu05a}). 

No magnetic field could be diagnosed in any Vega-like star.
The most accurate determination of a magnetic field, at 2.6\,$\sigma$ level, 
has been performed for the Vega-like star $\iota$\,Cen, for which we measured 
$\left<B_z\right>$=$-77\pm$30\,G.
In spite of expecting to find a magnetic field in $\beta$~Pictoris,
the prototype of Vega-like stars,
which shows conspicuous signs of chromospheric activity 
(Bouret et~al.\ \cite{Bou02}),
a longitudinal magnetic field is measured only
at $\sim$1.5\,$\sigma$ level.

A longitudinal magnetic field at a level higher than 3\,$\sigma$ was diagnosed
for the Herbig Ae stars HD\,31648, HD\,139614, and HD\,144432.
The main characteristics of the last two stars, HD\,139614 and HD\,144432,
were already 
discussed in our previous study (Hubrig et~al.\ \cite{Hu04b}). 
The pre-main sequence nature of HD\,31648 has been studied intensively at different 
wavelengths during the past years
since it was the first Herbig Ae star for which disk rotation was detected
(e.g., Mannings et~al.\ \cite{MKS97};
Eisner et~al.\ \cite{Eis04};
Vink et~al.\ \cite{Vin05}).
Imaging and interferometry studies have found an inclined
disk  with i$\sim$30$^\circ$ and PA$\sim$150$^\circ$.
The star is quite young with an age
of about 2.5--6.4 million years (van den Ancker et~al.\ \cite{Anc98}).
In the FORS\,1 low resolution spectra, the strong emission in the H$_\beta$ line is 
easily identified. The spectrum of this star in integral light is
shown in Fig.~\ref{fig:Stokes31648} (left).
Stokes V/I spectra of HD\,31648 over the whole
spectral region and in the spectral region around the \ion{Ca}{ii} doublet and nearby Balmer
lines are presented in Fig.~\ref{fig:Stokes31648} (centre and right).
The characteristic Stokes
V profiles indicative of the presence of a magnetic field 
are clearly detected in the \ion{Ca}{ii} H and K, and Balmer lines. 

\begin{figure}
\centering
\includegraphics[width=0.45\textwidth]{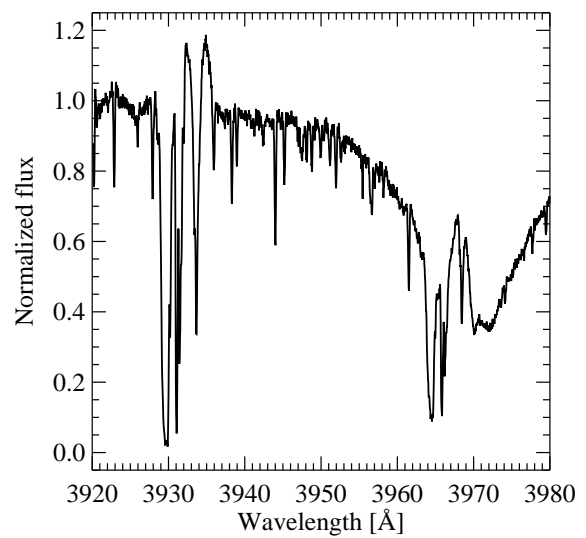}
\caption{
Observed profiles of \ion{Ca}{ii} H and K in the high resolution spectrum 
(R$\sim$48000) of HD\,190073 using the FEROS spectrograph
(Pogodin et~al.\ \cite{Pog05}).
\label{fig:ca}
}
\end{figure}

One Herbig Ae star in our sample, HD\,190073, was already measured in the past 
by Babcock (\cite{Bab58}) using Zeeman photographic spectra (see Sect.\,1). 
However, the presence of a magnetic field was not confirmed in the later study by
Glagolevskij \& Chountonov (\cite{GlCh01}).
This star, frequently called an unusual Ae star,
has a very low value of projected rotational velocity  $v\sin{i}$ and
displays some atypical characteristics along with properties common to other 
``classical'' Herbig stars. HD\,190073 appears to be much more luminous than
other Herbig stars of similar spectral type (van den Ancker et~al.\ \cite{Anc98}).
Quite recently, Pogodin et~al.\ (\cite{Pog05}) studied the
temporal behaviour of the H Balmer lines and the
\ion{He}{i} $\lambda$\,5876\,\AA{}, \ion{Na}{i}\,D,
and \ion{Ca}{ii} H and K lines in the spectrum of this star.
The most conspicuous spectral feature is the
complex structure of the \ion{Ca}{ii} H and K absorption profiles (Fig~\ref{fig:ca}).
It consists of a number of blueshifted absorption
components of different width and depth, and it demonstrates a complex
variability on timescales from months to decades.

In Fig.~\ref{fig:Stokes190073} (left) we present the low resolution
FORS\,1 spectrum (R$\sim$4000) of HD\,190073 
in the spectral region H$_\beta$ to H$_8$ to show the extraordinary rich emission line spectrum.
It is noteworthy that the Balmer lines are ``invisible'' in the Stokes V/I spectrum of
this star, and no Zeeman signature can be detected at their locations
(Fig.~\ref{fig:Stokes190073}, centre).
On the contrary, 
distinctive polarization signatures in the \ion{Ca}{ii} doublet appear exceptional by displaying
several components in both \ion{Ca}{ii} H and K lines
(Fig.~\ref{fig:Stokes190073}, right).
From the measurement of circular polarization in all Balmer lines from  H$_\beta$ to H$_8$,
we obtain $\left<B_z\right>$=$+26\pm$34\,G. 
However, using both \ion{Ca}{ii} H and K lines
for the measurement of circular polarization,
we are able to diagnose a longitudinal magnetic field at 
2.8\,$\sigma$ level, $\left<B_z\right>$=$+84\pm$30\,G.
This is the value presented in Table~\ref{tab:results}. 

Both \ion{Ca}{ii} H and K lines in the Stokes V/I spectrum of HD\,31648 display a similar
multicomponent complex structure. Since the rotational velocity of this star is rather 
high ($v\sin{i}=102$~km\,s$^{-1}$),  complex features in the \ion{Ca}{ii} H and K absorption 
profiles have not yet been reported by other studies of this star in integral light.
As mentioned in Sect.\,1, it is quite possible that a complex absorption profile structure 
of the \ion{Ca}{ii} lines is present in most Herbig Ae/Be stars and
might be caused by a specific topology of their magnetic fields.
However, due to the fast rotation of these stars,
these absorption components are smoothed out and remain undetected. 


\section{Discussion}\label{sec5}

The presented magnetic field measurements of Vega-like and Herbig Ae/Be stars 
demonstrate that longitudinal magnetic fields in these stars are rather weak.
The magnetic field in Vega-like
stars, if present at all, is less than 100\,G, and is almost 
below the detection limit of spectropolarimetric measurements with FORS\,1. 
A hint of a weak magnetic field is found in the Vega-like star HD\,115892, for which
the magnetic field is measured at the $\sim$2.6\,$\sigma$ level.
We should note that, as a matter of fact, the detection limit also 
depends on the structure of the magnetic
regions. In the present study we have almost reached the limits of what is 
currently feasible by using FORS\,1 in spectropolarimetric mode.

Confirmation of a magnetic field in the Herbig Ae star HD\,139614
and the new detections of the field in two other Herbig Ae stars, HD\,31648 and
HD\,144432, are encouraging and significant, as these detections confirm that the magnetic field
does play a role in the star-formation process and that the magnetic accretor scenario,
which is rapidly gaining popularity for intermediate-mass pre-main sequence stars,
may indeed explain how the PMS stars dissipate their angular momentum, keeping the star
rotating as slowly as is observed.
However, spectropolarimetric monitoring is required to enable us to probe the strength and 
geometry of the field responsible for the accretion. Within the magnetic accretor 
framework, we expect to see changes in polarization properties over the rotational 
cycle while spots and accretion columns cross the line of sight. Such spectropolarimetric 
monitoring observations have not been carried out yet.
The high accuracy measurements with FORS\,1 in spectropolarimetric mode presented here
are certainly motivations to ward the next step:
a monitoring program of one of the Herbig Ae/Be stars for which 
the diagnosed strength of the magnetic field is well above the FORS\,1 detection limit. 

As has been demonstrated for the Herbig Ae star HD\,190073
by Babcock (\cite{Bab58}), the magnetic field
can present different intensity and polarity for different elements and ions.
Accordingly, we would like to emphasise the importance of future magnetic field 
measurements
using high resolution spectropolarimeters similar to ESPaDOnS that was
recently installed at the CFHT (Manset \& Donati \cite{MD03}).
With such an instrument, it will be possible to measure
the magnetic field separately for lines of different elements in order to study both the magnetic
field configuration in Herbig Ae/Be stars and the interaction of the CS matter with the
magnetic field.

\begin{acknowledgements}
This research made use of the SIMBAD database,
operated at the CDS, Strasbourg, France.
\end{acknowledgements}

\end{document}